\documentclass[
reprint,
superscriptaddress,
 amsmath,amssymb,
 aps,
]{revtex4-1}

\usepackage{graphicx}
\usepackage{dcolumn}
\usepackage{bm}
\usepackage[mathlines]{lineno}
\usepackage[colorlinks=true]{hyperref}
\usepackage{mathrsfs}
\usepackage[usenames, dvipsnames]{xcolor}
\usepackage{float}
\usepackage{ulem}
\usepackage[utf8]{inputenc}
\usepackage{csquotes}

\def\be{\begin{equation}}
\def\ee{\end{equation}}
\def\ba{\begin{eqnarray}}
\def\ea{\end{eqnarray}}

\def\nn{\nonumber}

               \def\l {\lambda}   \def\c {\chi}       
              \def\.{\cdot}

\begin{document}

\title{Escape probability of particle from Kerr-Sen black hole}

\author{Ming Zhang}
\email{mingzhang@jxnu.edu.cn}
\affiliation{Department of Physics, Jiangxi Normal University, Nanchang 330022, China}
\author{Jie Jiang}
\email{Corresponding author, jiejiang@mail.bnu.edu.cn}
\affiliation{Department of Physics, Jiangxi Normal University, Nanchang 330022, China}
\affiliation{Department of Physics, Beijing Normal University, Beijing 100875, China}

\begin{abstract}
Assuming that a particle source is at rest in a locally non-rotating frame on the equatorial plane of the Kerr-Sen black hole,  we investigate the escape of the massless particle and massive particle from the black hole to spatial infinity. We calculate the escape probabilities of the emitted particles. We find that the angular momentum of the Kerr-Sen black hole boosts the escape probabilities near the horizon of the extreme Kerr-Sen black hole; however the angular momentum of the Kerr-Sen black hole suppresses the escape probabilities near the horizon of the non-extreme Kerr-Sen black hole. We also discover that the horizon limit of the Kerr-Sen black hole is not the most difficult position to escape for a particle in certain conditions.

\end{abstract}

\maketitle

\section{Introduction}
The Penrose process \cite{Penrose:1971uk}, which is well-known as an energy extraction process from ergosphere of the rotating black hole by one emitted particle originating from an initial ingoing one which breaks up, has been developed into the collisional Penrose process \cite{piran1975high} for astrophysical considerations, as it was pointed out that the process is  unplausible in astrophysics \cite{Bardeen:1972fi}. Generally, there are three kinds of collisional Penrose process. The first case \cite{Piran:1977dm} is that two (massive or massless) particles fall into the ergosphere of the rotating black hole from spatial infinity at rest and then collide with each other, producing a particle that can escape to infinity. The second case \cite{Schnittman:2014zsa,Mukherjee:2018kju} is that one particle with an impact parameter greater than the critical value falls into the black hole from spatial infinity and then turns around, colliding with the other ingoing particle. The third case, which is named as the super Penrose process \cite{Berti:2014lva,Zaslavskii:2014jea,Zhang:2018gpn}, is that two particles take a head-on collision in the ergosphere of the black hole. In this process, one of the particles comes from infinity and the other is a consequence of previous scattering events (for more discussions on the origination of this outgoing particle, see Refs. \cite{Zaslavskii:2017guu,Leiderschneider:2015ika,Zaslavskii:2015ema,Zaslavskii:2015fqy}). The efficiencies of these processes are quite different, as
\begin{equation}
\eta_{\text{penrose}}\approx 121\% <\eta_{\text{first}}<\eta_{\text{second}}<\eta_{\text{third}}\to \infty.
\end{equation}
Besides the (Collisional) Penrose process, another high-energy event is the Bañados-Silk-West mechanism, which states that the center-of-mass energy of two test particles can be arbitrarily high in the condition that the collision between them happens near the event horizon of the extremal black hole and the angular momentum and the energy of one particle follow a critical relation \cite{Banados:2009pr,Sadeghi:2011qu,Sadeghi:2013gmf,Pourhassan:2015lfa,An:2017hby,Guo:2016vbt,Jiang:2019cuc}. What is more, the observation of  the shadow \cite{Akiyama:2019cqa,Akiyama:2019fyp,Akiyama:2019eap} of the black hole depends on the photons that escape from the black hole.

The observations of the high-energy astrophysical events and the black hole shadow depend on the massive particle and photons we can observe. Of particular interest is the escape probability of the particle in the strong gravitational field. Particularly, the escape probability of the photon  determines the brightness of the halo around the black hole we observe. The escape probability of the produced particles in the super Penrose process for the extreme Kerr black hole was studied in \cite{Ogasawara:2016yfk}. It was shown that the resulting massless particles with extremely high energy in the super Penrose process can escape to infinity with a probability of $5/12$ and an impact parameter close to $2M$, where $M$ is the mass of the Kerr black hole. It motivates our research in this paper. We are interested in not only the escape probability of the  massless photon but also that of the massive particle. We will study the observability of the massless and massive particles emitted from the Kerr-Sen black hole. 

The remaining parts of this paper are arranged as follows. In section \ref{s2}, we will show the four-momentum of a particle in the Kerr-Sen spacetime using a locally non-rotating frame (LNRF). In section \ref{10009}, we will calculate the escaping probability of the massless particle from the extreme Kerr-Sen black hole. In section \ref{10010}, we will evaluate the escape probability of the massive particle from the extreme Kerr-Sen black hole. In section \ref{10011}, we will extend our study of the emission of the massless and massive particle from the extreme Kerr-Sen black hole to the non-extreme Kerr-Sen black hole case. Section \ref{10013} will be employed to discuss the escape probability variation of the particle with respect to the position of the particle source.   Section \ref{5411} will be devoted to our conclusions and discussions.

\section{Four-momentum of the particle in the Kerr-Sen spacetime}\label{s2}
The line element of the Kerr-Sen black hole is \cite{Sen:1992ua}
\begin{equation}
\begin{aligned}
ds^2=&-\frac{\Delta-a^2\sin^2\theta}{\Sigma}dt^2+\frac{\Sigma}{ \Delta} dr^2+\Sigma d \theta^2\\&+\frac{\Xi \sin^2\theta}{\Sigma}d\phi^2-\frac{4Mra\sin^2 \theta}{\Sigma}dtd\phi,
\end{aligned}
\end{equation}
where
\begin{align}
\Sigma&=r(r+2c)+a^2\cos^2\theta,\nn\\ \Delta&=r(r+2c)-2Mr+a^2,\nn\\ \Xi&=[a^2+r(2c+r)]^2-\Delta a^2\sin^2\theta,\nn\\c&=Q^2/2M\nn.
\end{align}
The inner and outer horizons of the black hole are given as
\begin{align}
r_\text{i}&=M-c-\sqrt{\left(M-c\right)^2-a^2},\\ r_\text{h}&=M-c+\sqrt{\left(M-c\right)^2-a^2}.
\end{align}
When $|a| = \left|M-c\right|$, the Kerr-Sen black hole becomes extreme.

There are two Killing vectors, $\xi^{t}\equiv \delta^{t}_{\mu}$ and $\xi^{\phi}\equiv \delta^{\phi}_{\mu}$, for the stationary Kerr-Sen black hole. Correspondingly, the conserved quantities, namely conserved energy and angular momentum of a particle, read
\begin{align}
e&=-g_{t\nu}\xi^{t} p^\nu, \\l&=g_{\phi\nu}\xi^{\phi} p^\nu,
\end{align}
where $p^{\mu}$ is the four-momentum of the particle with mass $m$. $p^{\mu}p_{\mu}=-m^{2}$, $m=0$ for the massless particle (photon) and $m>0$ for the massive particle. Not loss of generality, by setting $\theta=\pi/2$, meaning that the particle moves on the equatorial plane, we obtain the particle's four-momentum
 \begin{align}
p^t&=\frac{2 a l M r \Sigma -e \Xi  \Sigma }{a^2 \left(\Xi -4 M^2 r^2\right)-\Delta  \Xi },\\
p^r&=\sigma\sqrt{V(r)},\label{effall}\\
p^\theta&=0,\label{ptheta}\\
p^\phi&=\frac{\Sigma  \left(-a^2 l+2 a e M r+\Delta  l\right)}{\Delta  \Xi -a^2 \left(\Xi -4 M^2 r^2\right)},
\end{align}
where $\sigma=\pm 1$ correspond respectively to the particles that move radially outwards and inwards,
\begin{equation}
V(r)=\frac{-a^2 \Delta  p_{t}^2+4 a \Delta  M p_{t} p_{\phi} r-\Delta  m^2 \Sigma +\Delta ^2 p_{t}^2-\Delta  \Xi  p_{\phi}^2}{\Sigma ^2}
\end{equation} 
can be viewed as the radial effective potential of the particle.

To describe the particle moving in the rotating Kerr-Sen spacetime, we resort to an orthonormal tetrad accompanying with a locally non-rotating observer, namely LNRF \cite{Bardeen:1972fi}, which can be written as
\begin{equation}
\begin{aligned}
e^{(t)}_\mu&=\left(\sqrt{\frac{\Delta  \Sigma }{\Xi }},0,0,0\right),\\
e^{(r)}_\mu&=\left(0,\sqrt{\frac{\Sigma }{\Delta }},0,0\right),\\
e^{(\theta)}_\mu&=(0,0,\sqrt{\Sigma },0),\\
e^{(\phi)}_\mu&=\left(-\frac{2 a M r}{\sqrt{\Xi  \Sigma }},0,0,\sqrt{\frac{\Xi }{\Sigma }}\right).
\end{aligned}
\end{equation}
The tetrad basis is related to the metric by the transformation
\begin{equation}
g_{\mu\nu}=\eta_{(a)(b)}e^{(a)}_{\mu}e^{(b)}_{\nu},\quad \eta_{(a)(b)}=\text{diag}(-1,1,1,1).
\end{equation}
Then by the translation
\begin{equation}
p^{(\alpha)}=e^{(\alpha)}_\mu p^\mu,
\end{equation}
we can obtain the four-momentum of the particle in the LNRF.

\begin{figure}[!htbp] 
   \centering
   \includegraphics[width=2.25in]{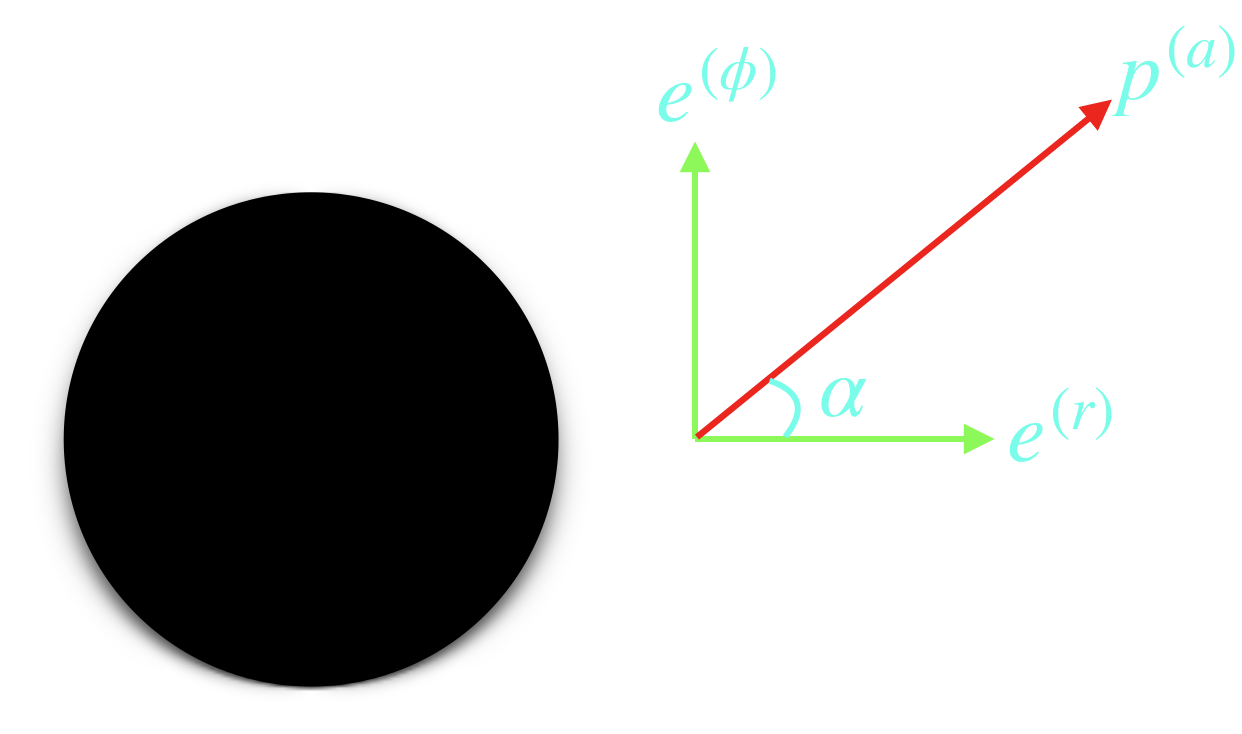}
   \caption{Emission angle of the particle in the LNRF. $p^{(a)}$ is the four-momentum of the particle.}
   \label{angle}
\end{figure}

\section{Escaping probability of the massless particle from extreme Kerr-Sen black hole at horizon limit}\label{10009}
We will now evaluate the escaping probability of the massless particle from the extreme Kerr-Sen black hole produced by a source at rest in LNRF at the horizon limit. The four-momentum of the massless particle in the LNRF can be obtained as
 \begin{align}
p^{(t)}=&\frac{e Z \left(2 M (M+r)+Q^2\right)+4 l M^3 \left(Q^2-2 M^2\right)}{\left(2 M (r-M)+Q^2\right) \sqrt{Z \left(M r+Q^2\right)}},\\
p^{(r)}=&\frac{2\sigma\sqrt{V(r)} \sqrt{M r \left(M r+Q^2\right)}}{-2 M^2+2 M r+Q^2},\\
p^{(\theta)}=&0,\label{ptheta}\\
p^{(\phi)}=&\frac{2 l M \left(M r+Q^2\right)}{\sqrt{Z \left(M r+Q^2\right)}},
\end{align}
where
\begin{equation}
\begin{aligned}
Z=&\left[4 M^4-2 M^3 r+2 M^2 \left(r^2-2 Q^2\right)+3 M Q^2 r+Q^4\right]\\&\times \left(2 M (M+r)+Q^2\right)\nn.
\end{aligned}
\end{equation}

We consider a particle source at rest in the LNRF on the equatorial plane of the Kerr-Sen black hole at the position $r_{*}$. We can introduce the emission angle $\alpha$ with respect to the LNRF tetrad (see Fig. \ref{angle}), which can be determined by the relations among the components of $p^{(a)}$ for the massless particle, as
\begin{equation}
\begin{aligned}
\sin\alpha=&
\frac{p^{(\phi)}}{\sqrt{\left(p^{(r)}\right)^2+\left(p^{(\phi)}\right)^2}}=\frac{p^{(\phi)}}{p^{(t)}}\\
=&\frac{2 b M \left(M r_{*}+Q^2\right) \left(2 M (r_{*}-M)+Q^2\right)}{4 b M^3 \left(Q^2-2 M^2\right)+Z(r_*)},
\end{aligned}
\end{equation}
\begin{equation}
\begin{aligned}
\cos\alpha=&
\frac{p^{(r)}}{\sqrt{\left(p^{(r)}\right)^2+\left(p^{(\phi)}\right)^2}}=\frac{p^{(r)}}{p^{(t)}}\\
=&\frac{\sqrt{\xi } \sqrt{Z(r_*)} \left(2 M (M-r_{*})-Q^2\right)}{\sqrt{M} \left(4 b M^3 \left(Q^2-2 M^2\right)+Z(r_*)\right)},
\end{aligned}
\end{equation}
where
\begin{equation}
\begin{aligned}
\xi =&b^2 M \left(2 M^2-M r_{*}-Q^2\right)+2 b M^2 \left(Q^2-2 M^2\right)\\&+M \left(M-\frac{Q^2}{2 M}\right)^2 \left[M (2 M+r_{*})+Q^2\right]\\&+r_{*}\left(M r_{*}+Q^2\right)^2\nn,
\end{aligned}
\end{equation}
with the impact parameter defined by $b\equiv l/e$.

\begin{figure}[!htbp] 
   \centering
   \includegraphics[width=2.45in]{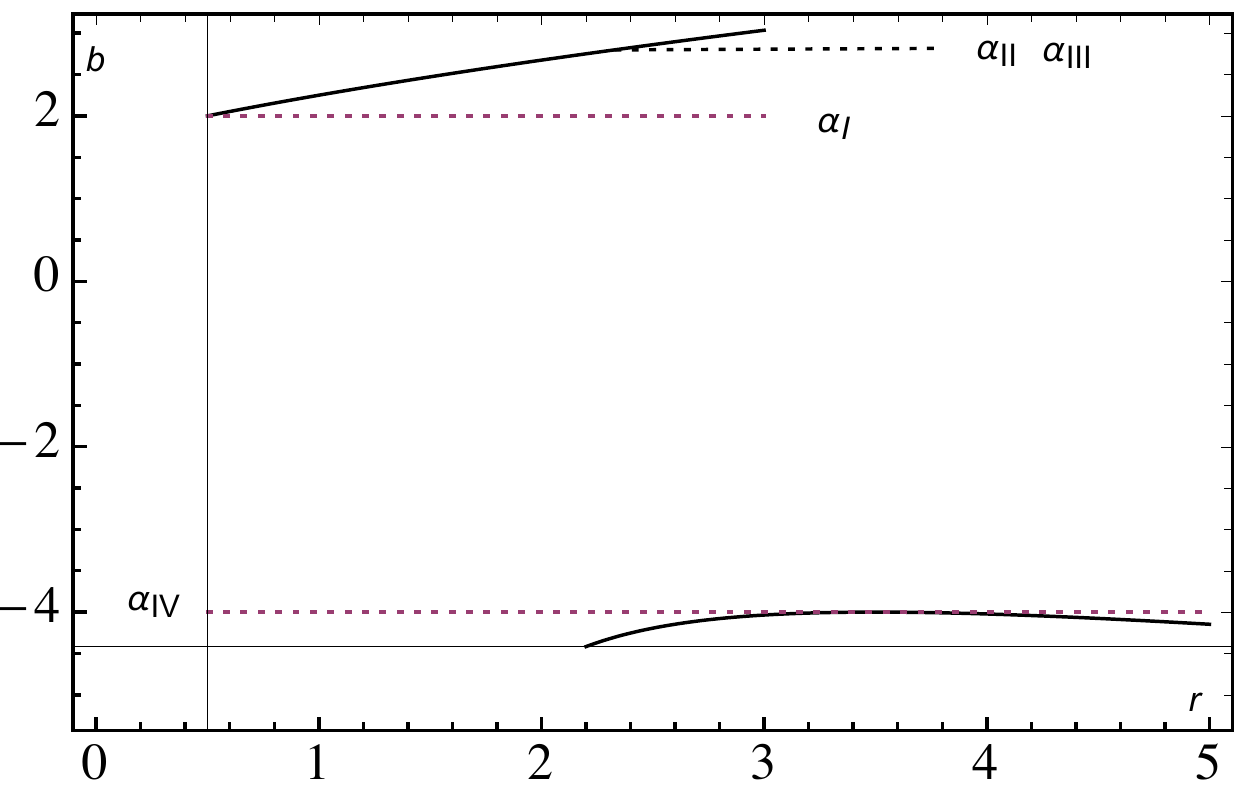}
   \caption{A schematic diagram of the impact parameter $b$ of the particle.}
   \label{sch}
\end{figure}

Letting $V(r)=0$, we can obtain the impact parameters 
\begin{eqnarray}
b_+(r)&:=&\frac{Q^2}{2 M}+M+r,\\
b_-(r)&:=&-\frac{2 M \left(2 M^2-Q^2\right)}{M (r-2 M)+Q^2}-\frac{Q^2}{2 M}-M-r.
\end{eqnarray}
Then we can know particles with the impact parameter
\begin{equation}
 2 M<b<b_+(r_*)
 \end{equation}
can escape from the extreme Kerr-Sen black hole to spatial infinity for any sign of initial radial velocity; for the particle with the impact parameter 
\begin{equation}
b_{e}<b<2M,
\end{equation}
with
\begin{equation}
b_{e}=\frac{-4 M \sqrt{4 M^2-2 Q^2}-6 M^2+Q^2}{2 M}
\end{equation}
the extreme point of $b_{-}(r)$, it cannot escape to spatial infinity if its initial radial velocity is inward. Then according to the schematic diagram in Fig. \ref{sch}, critical angles related to the critical impact parameters that the massless particle can escape to infinity  can be defined as
\begin{align}
\alpha_{\rm I}
&\equiv\alpha[\sigma_3=-1,\;b=2M],\\
\alpha_{\rm II}
&\equiv\alpha[\sigma_3=-1,\;b=b_+(r_*)],\\
\alpha_{\rm I\hspace{-.1em}I\hspace{-.1em}I}
&\equiv\alpha[\sigma_3=1,\;b=b_+(r_*)],\\
\alpha_{\rm IV}
&\equiv\alpha[\sigma_3=1,\;b=b_{e}].
\end{align}
As 
\begin{equation}
\sin\alpha_{\rm II}=\sin\alpha_{\rm I\hspace{-.1em}I\hspace{-.1em}I}=1\nn
\end{equation}
and 
\begin{eqnarray}
\sin\alpha_{\rm I}>0,\;\cos\alpha_{\rm I}<0,\;\sin\alpha_{\rm I\hspace{-.1em}V}<0,\;\cos\alpha_{\rm I\hspace{-.1em}V}>0,\nn
\end{eqnarray}
we have
\begin{equation}
\alpha_{\rm I\hspace{-.1em}V}<\alpha_{\rm I\hspace{-.1em}I\hspace{-.1em}I}=\alpha_{\rm II}=\frac{\pi}{2}<\alpha_{\rm I},\nn
\end{equation}
which indicates that the massless particle that escapes from the extreme Kerr-Sen black hole to spatial infinity must has an angle
\begin{equation}
\alpha\in(\alpha_{\rm I\hspace{-.1em}V},\alpha_{\rm I}).
\end{equation}

To obtain concrete values of the characterized critical angles, we can set that the collision takes place at a near-horizon point of the extreme Kerr-Sen black hole
\begin{equation}
r_*=\frac{M-Q^2/(2M)}{1-\epsilon},\nn
\label{r_collision}
\end{equation}
with $0<\epsilon\ll1$.
Then the critical angles can be directly calculated as
\begin{align}
\alpha_{\rm I}&=\pi-\arcsin\left(\frac{2 M^2+Q^2}{4 M^2}\right)+\mathcal{O}(\epsilon^0 ),\label{sol_alpha1}\\
\alpha_{\rm I\hspace{-.1em}I}&=\alpha_{\rm I\hspace{-.1em}I\hspace{-.1em}I}=\frac{\pi}{2},\label{sol_alpha23}\\
\alpha_{\rm I\hspace{-.1em}V}&=-\frac{\epsilon  \left(2 M^2+Q^2\right) \varsigma}{4 M^2 \left(4 M^3+\varsigma\right)}+\mathcal{O}(\epsilon),
\end{align}
where
\begin{equation}
 \varsigma=6 M^3+4 \sqrt{4 M^6-2 M^4 Q^2}-M Q^2.\nonumber
 \end{equation}
When $Q\to \sqrt{2}M$, we have
\begin{align}
\alpha_{\rm I}=\frac{\pi}{2},\quad
\alpha_{\rm I\hspace{-.1em}V}=-\frac{\epsilon }{2}.
\end{align}

The escape probability depends on the escape cone from which the photon can escape to spatial infinity. If viewing that the particle on the equatorial plane emitted isotropically, we can calculate the escape probability of the massless particle produced by the particle source near the event horizon of the extreme Kerr-Sen black hole by comparing the escape cone encircled by $\alpha_{i}$ ($i={\rm I, II, III, IV}$) with $2\pi$ \cite{Ogasawara:2016yfk},
\begin{equation}\label{806282}
\begin{aligned}
P\equiv\frac{\alpha_{\rm I}-\alpha_{\rm I\hspace{-.1em}V}}{2\pi},
\end{aligned}
\end{equation}
which gives escape probability of the massless particle as
\begin{equation}\label{806280}
\begin{aligned}
P=\frac{1}{2}-\frac{1}{2\pi}\arcsin\left(\frac{1}{2}+\frac{Q^2}{4 M^2}\right).
\end{aligned}
\end{equation}
It is evident that the escape probability of the particle decreases with the increasing electric charge of the extreme Kerr-Sen black hole. Especially, in the limits $\epsilon\to0, Q\to \sqrt{2}M$, the escape probability becomes
\begin{equation}
P\to\frac{1}{4},
\end{equation}
and for $Q\to 0$, we have $P\to 5/12$, which is just the result of extreme Kerr black hole obtained in \cite{Ogasawara:2016yfk}.

\section{Escaping probability of massive particle from extreme Kerr-Sen black hole at horizon limit}\label{10010}
For convenience of calculations, we set $M=m=1$ for what follows in this passage. For the massive particle, we can introduce the emission angle $\beta$ from the particle source with respect to the LNRF tetrad, which can be determined by the relations among the components of $p^{(a)}$ for the massive particle as
\begin{equation}
\begin{aligned}
\sin\beta=\frac{p^{(\phi)}}{\sqrt{\left(p^{(r)}\right)^2+\left(p^{(\phi)}\right)^2}},
\end{aligned}
\end{equation}
\begin{equation}
\begin{aligned}
\cos\beta=\frac{p^{(r)}}{\sqrt{\left(p^{(r)}\right)^2+\left(p^{(\phi)}\right)^2}}.
\end{aligned}
\end{equation}
$V(r)=0$ gives the impact parameters
\be
\bar{b}_{+}(r)=\frac{2  Q^2-4 +\sqrt{\mathcal{X}_{1} \left(Q^2+r\right)} \left(Q^2+2 r-2\right)/e}{2 \left(Q^2+r-2\right)},
\ee
\be
\bar{b}_{-}(r)=\frac{2  Q^2-4 -\sqrt{\mathcal{X}_{1} \left(Q^2+r\right)} \left(Q^2+2 r-2\right)/e}{2 \left(Q^2+r-2\right)},
\ee
where $\mathcal{X}_{1}=e^2 \left(Q^2+r\right)- \left(Q^2+r-2\right)$. Letting $r\to r_{h}$ for the extreme Kerr-Sen black hole, we can obtain
$\bar{b}_{+}\to 2$. Denoting the extreme value of $\bar{b}_{-}$ as $\bar{b}_{e}$, we can know that  for $ 2 <\bar{b}<\bar{b}_+(r_*)$, the massive particle can escape from the extreme Kerr-Sen black hole to spatial infinity irrespective of the initial sign of its radial velocity and for $\bar{b}_{e}<\bar{b}<2$, the escape can only happen for the particle with outward initial radial velocity.

\begin{table*}[!htb]
\centering
\caption{The critical escape angle $\beta_{I}$ for a massive particle from the extreme Kerr-Sen black hole at the event horizon limit.}
\begin{tabular}{cccccccccc} 
  \hline\hline
$Q~~~$ &~~~0~~~&~~~0.1~~~&~~~0.3~~~&~~~0.5~~~&~~~0.7~~~&~~~0.9~~~&~~~1.1~~~&~~~1.3~~~&~~~$\sqrt{2}$\\
$\sin\beta_{I}~~~$ &~~~$1/\sqrt{3}$~~~&~~~0.58~~~&~~~0.60~~~&~~~0.65~~~&~~~0.71~~~&~~~0.79~~~&~~~0.87~~~&~~~0.96~~~&~~~1\\
$\cos\beta_{I}~~~$ &~~~$-\sqrt{2/3}$~~~&~~~-0.81~~~&~~~-0.80~~~&~~~-0.76~~~&~~~-0.70~~~&~~~-0.61~~~&~~~-0.48~~~&~~~-0.29~~~&~~~0\\
\hline
\hline
\end{tabular}
  \label{table1}
\end{table*}

\begin{figure}[!htbp] 
   \centering
   \includegraphics[width=2.5in]{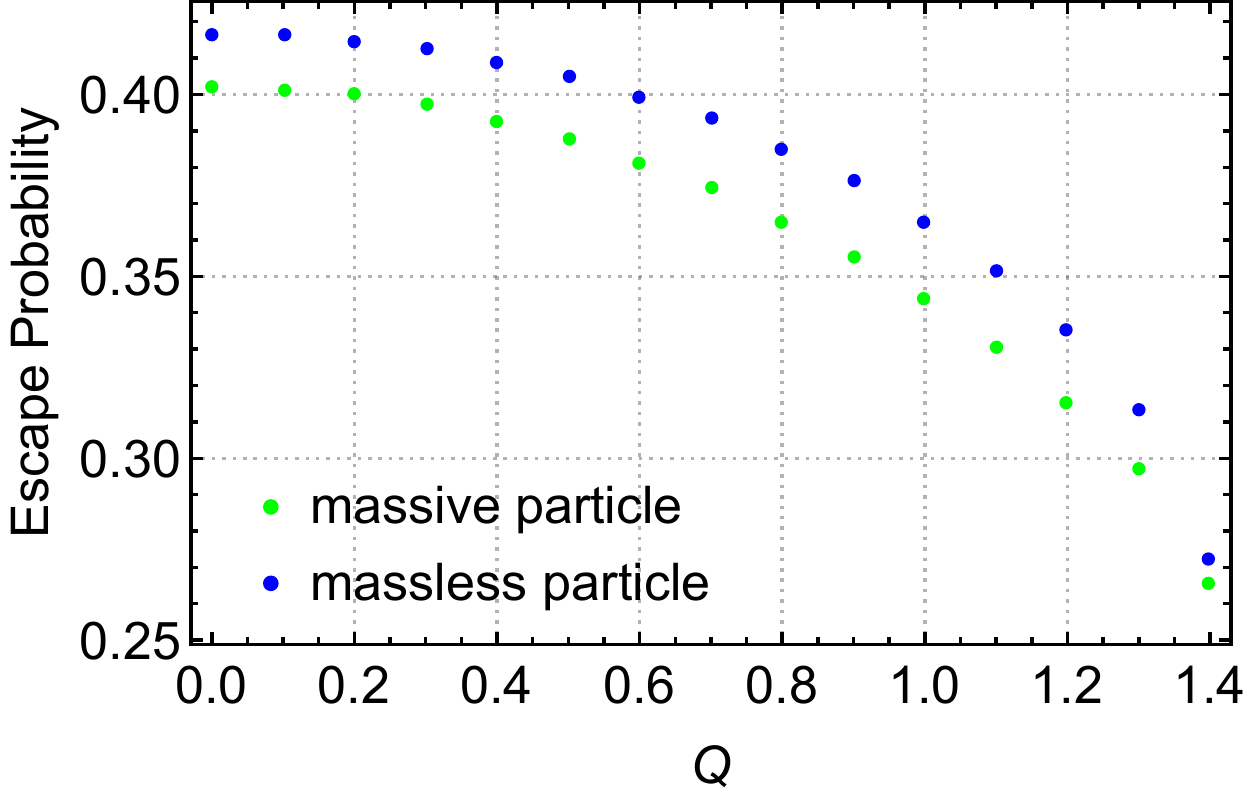}
   \caption{Escape probabilities of the massive and massless particles from the extreme Kerr-Sen black hole at the event horizon limit with $M=1$.}
   \label{ep}
\end{figure}

Like the massless case, the critical angles related to the critical impact parameters that the massive particle can escape to infinity  can be defined as
\begin{align}
\beta_{\rm I}
&\equiv\beta[\sigma_3=-1,\;\bar{b}=2],\\
\beta_{\rm II}
&\equiv\beta[\sigma_3=-1,\;\bar{b}=\bar{b}_+(r_*)],\\
\beta_{\rm I\hspace{-.1em}I\hspace{-.1em}I}
&\equiv\beta[\sigma_3=1,\;\bar{b}=\bar{b}_+(r_*)],\\
\beta_{\rm IV}
&\equiv\beta[\sigma_3=1,\;\bar{b}=\bar{b}_{e}].
\end{align}
Setting the conserved energy of the massive particle as $e=1$ (so that the particle can escape from the black hole just to an observer at spatial infinity) gives
\begin{align}
\sin\beta_{\rm II}&=\sin\beta_{\rm I\hspace{-.1em}I\hspace{-.1em}I}=1,\\ \cos\beta_{\rm II}&=\cos\beta_{\rm I\hspace{-.1em}I\hspace{-.1em}I}=0,
\end{align}
which means
\begin{equation}
\beta_{\rm II}=\beta_{\rm I\hspace{-.1em}I\hspace{-.1em}I}=\frac{\pi}{2}.
\end{equation}
The critical compact parameter of $\bar{b}_{-}$ is
\be
\bar{b}_{e}=\frac{-2 \sqrt{2-Q^2} \left(Q^2+\mathcal{X}_2\right)+Q^2-2}{Q^2+\mathcal{X}_2-2},
\ee
where 
\be
\mathcal{X}_{2}=\frac{1}{2} \left(-3 Q^2+4 \sqrt{2-Q^2}+6\right).\nn
\ee
Then we have
\be
\sin\beta_{\text{IV}}=0,\,\cos\beta_{\text{IV}}=1,
\ee
which gives 
\be
\beta_{\text{IV}}=0.
\ee
The critical angles $\beta_{I}$ can be obtained numerically and we have listed them in Table \ref{table1}. The escape probability of the massive particle in the extreme Kerr-Sen black hole spacetime background is
\be\label{massiveep}
\bar{P}=\frac{\beta_{\text{I}}-\beta_{\text{IV}}}{2\pi}.
\ee
We show the escape probability of the massive particle together with that of the massless particle at the event horizon limit of the extreme Kerr-Sen black hole in Fig. \ref{ep} according to Eqs. (\ref{806280}) and (\ref{massiveep}). From the figure, we can see that, for both the two kinds of particles, the escape probabilities decrease with the increasing electric charge of the extreme Kerr-Sen black hole. This signifies that the rotation of the black hole promotes the escape of both the massless and massive particles. Besides, we can see that the escape probability of the massless particle is always greater than that of the massive particle at the event horizon limit. The two branches intersect at $Q=\sqrt{2}$ where the angular momentum of the extreme black hole vanishes. At that point, both of the two kinds of particles own an escaping probability 1/4. 

\begin{figure}[!htbp] 
   \centering
   \includegraphics[width=2.5in]{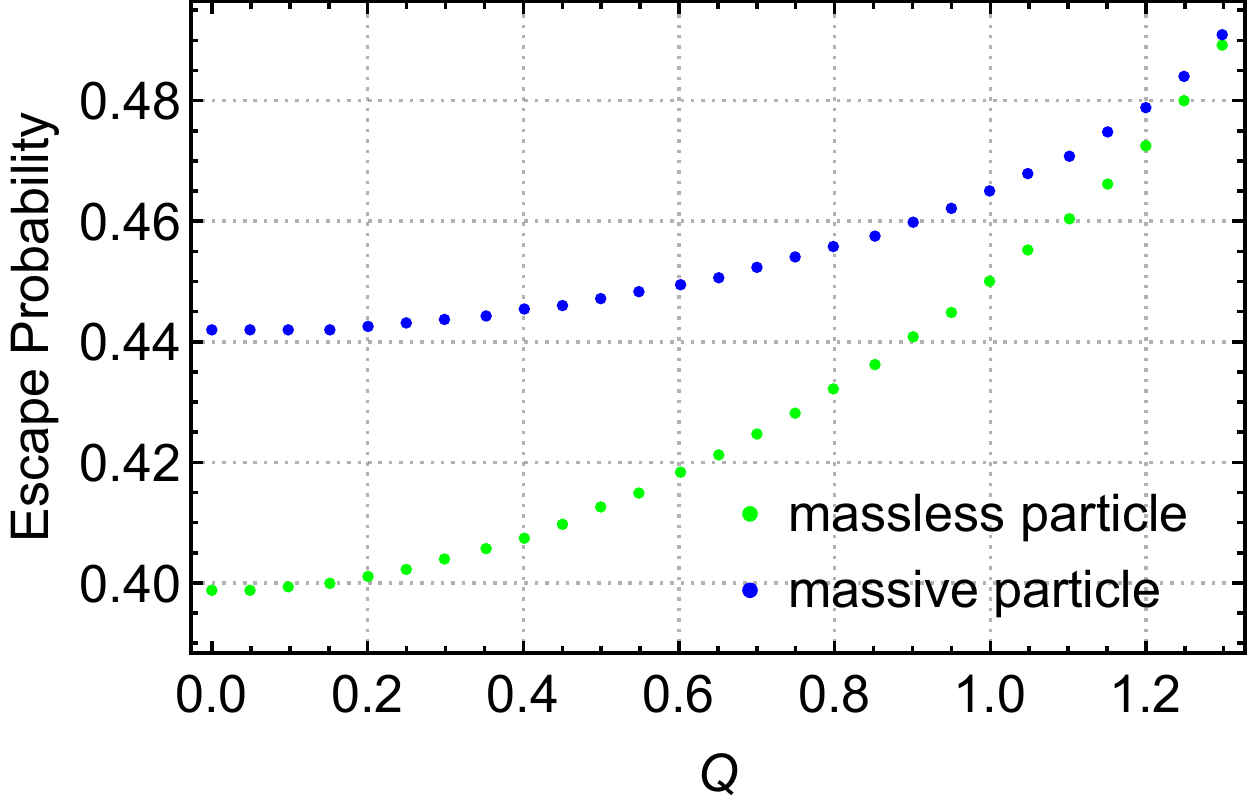}
   \caption{Escape probabilities of the massive and massless particles from the non-extreme Kerr-Sen black hole at the domain of outer communication region with $K=0.9$. We choose the position of the particle source at $r_{*}=4$. }
   \label{ep2}
\end{figure}

\begin{figure}[!htbp] 
   \centering
   \includegraphics[width=2.9in]{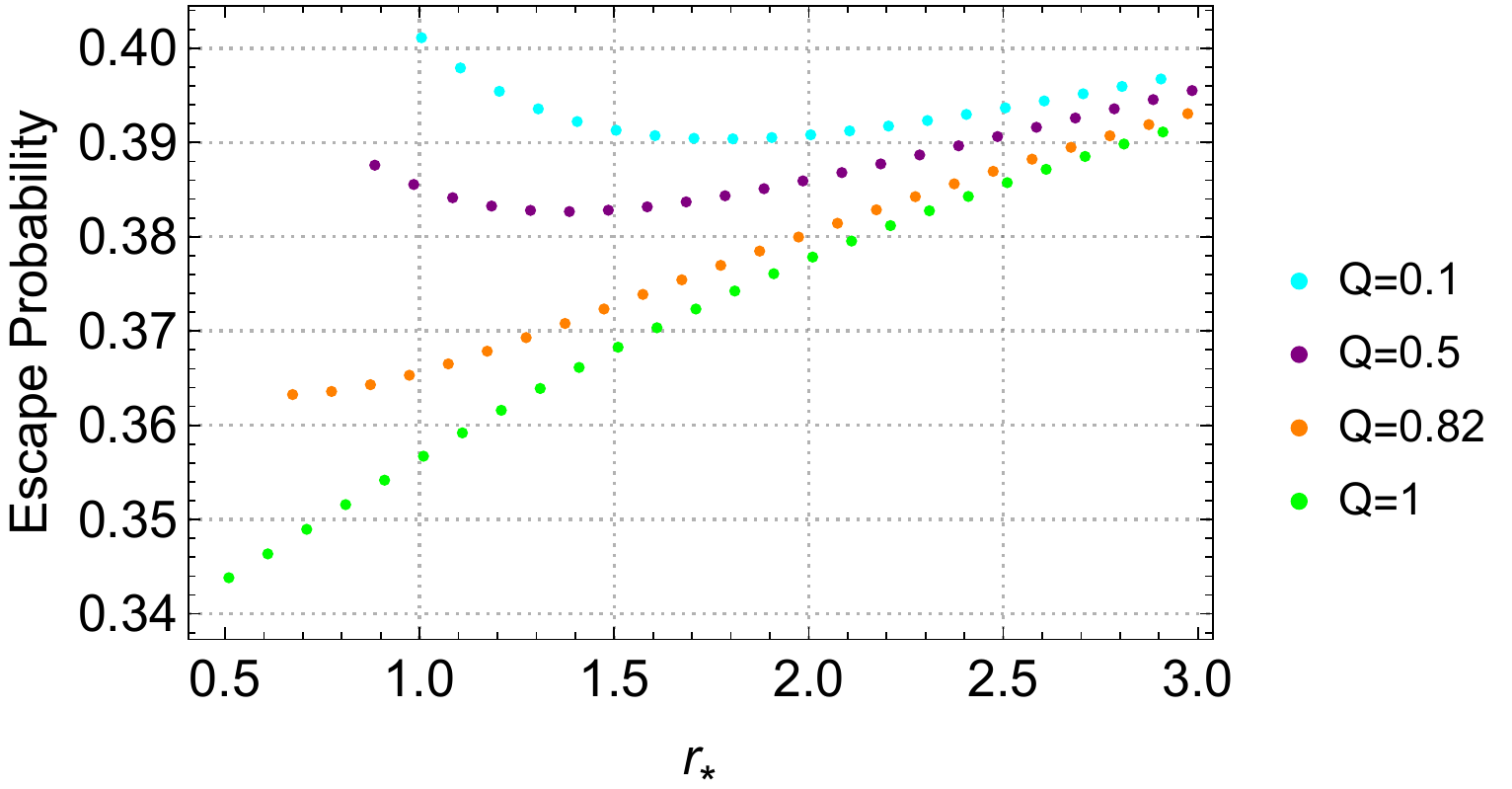}\\
      \includegraphics[width=2.9in]{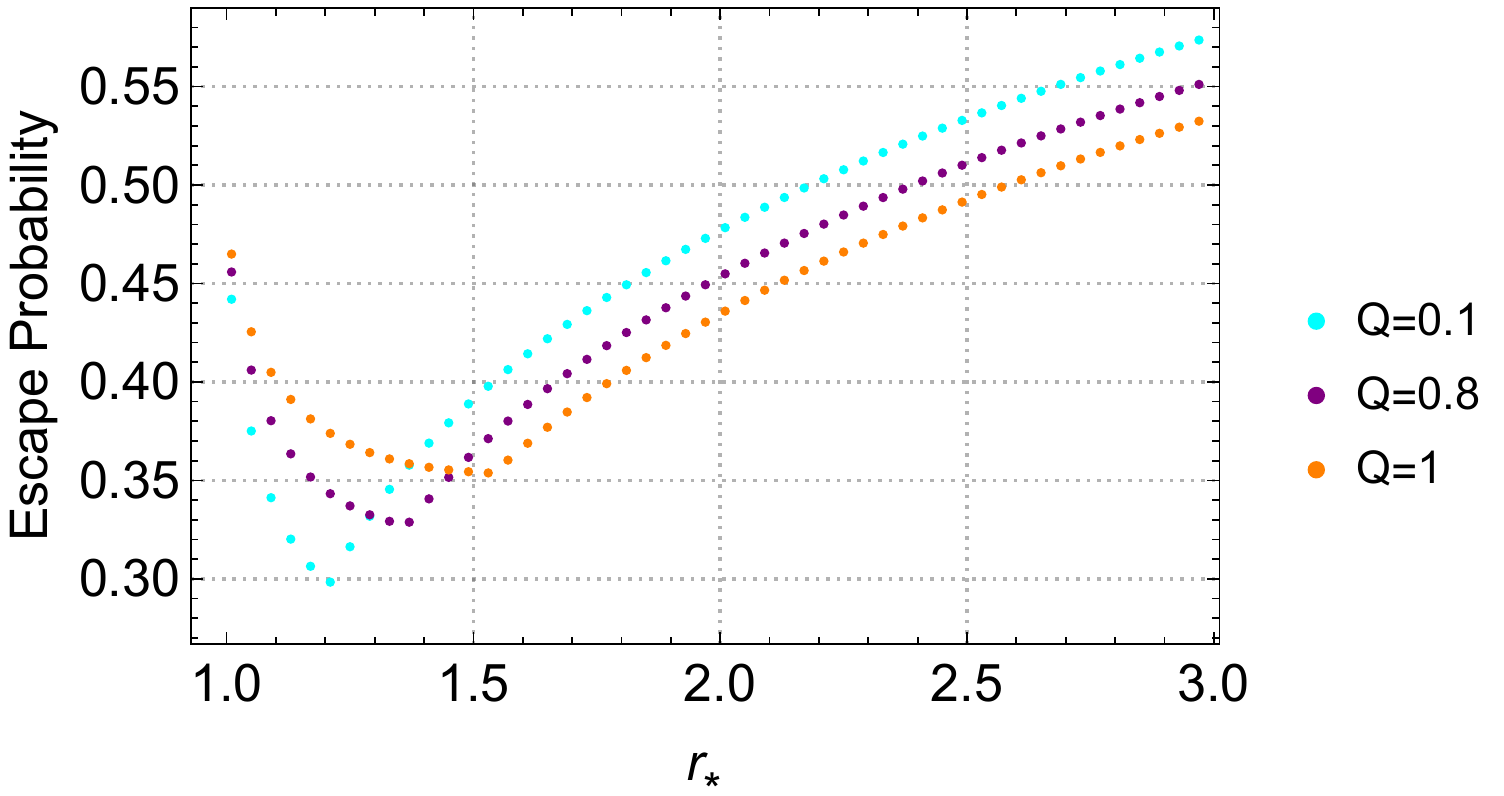}
   \caption{Top panel: variations of the escape probabilities of the massive particle in the domain of outer communication region of extreme Kerr-Sen black hole with respect to the position of the particle source. Bottom panel: variations of the escape probabilities of the massive particle in the domain of outer communication region of the non-extreme Kerr-Sen black hole with respect to the position of the particle source for K=0.9.}
   \label{ep3}
\end{figure}
 
 \section{Escape probabilities of  particles from a source near the horizon of the non-extreme Kerr-Sen black hole}\label{10011}
 We will probe the escape probabilities of the massless and massive particles escaping from a particle source near the horizon of  the non-extreme Kerr-Sen black hole. The regularity condition of the non-extreme Kerr-Sen black hole gives $a+Q^{2}/2<1$. We now investigate the escape probability of the massless and massive particle from the non-extreme Kerr-Sen black hole, with $a+Q^{2}/2=K<1$ and $a\geqslant 0$ (so that $Q\leqslant \sqrt{2K}$). We will choose $K=0.9$ for instance.

In the previous extreme cases for the massless and massive particles, the effective potential $V(r)$ in (\ref{effall}) is fulfilled at and near the event horizon. In the non-extreme case, we should care about the demand of $V(r)\geqslant 0$. We will not consider the event horizon limit as $V(r)< 0$ there; instead, we will choose a position where the effective potentials of the massless and massive particles are fulfilled.

It is evident that the impact parameters at the position $r_{*}$ for the particles make their critical escape angles to be
\be
\alpha_{\rm II}=\alpha_{\rm I\hspace{-.1em}I\hspace{-.1em}I}=\beta_{\rm II}=\beta_{\rm I\hspace{-.1em}I\hspace{-.1em}I}=\frac{\pi}{2}.
\ee
For the extreme Kerr-Sen black hole, we have $\beta_{\text{IV}}\sim\bar{\beta}_{\text{IV}}\sim 0$ (even not at the event horizon limit). However, for the non-extreme Kerr-Sen black hole, this is not the case. For $K=0.9$,
\be
-\frac{\pi}{2}<\beta_{\text{IV}}<0,\quad -\frac{\pi}{2}<\bar{\beta}_{\text{IV}}<0.
\ee
When $Q\to 0$, $\beta_{\text{IV}}\sim \bar{\beta}_{\text{IV}}\to -\pi/2$.

For both of the two kinds of particles, we can obtain the same impact parameter at the event horizon of the non-extreme Kerr-Sen black hole as
\be
b_{+}=\bar{b}_{+}=\frac{2 \left(Q^2-2 K\right)}{2 \sqrt{(K-1) \left(-K+Q^2-1\right)}+Q^2-2},
\ee
with which we have
\be
\frac{\pi}{2}<\beta_{\text{I}}<\pi,\quad \frac{\pi}{2}<\bar{\beta}_{\text{I}}<\pi,
\ee
for $K=0.9$. When $Q\to \sqrt{2K}$, $\beta_{\text{I}}\sim \bar{\beta}_{\text{I}}\to \pi/2$.

We show the escape probabilities for both of the two kinds of particles in Fig. \ref{ep2}. (Note that one branch of the impact parameters for the particle around the non-extreme Kerr-Sen black hole at the horizon is not minimal and the minimal value can be got at a radial position beyond the horizon. We have considered this fact when we calculate the escape probability.) We can see that the escape probabilities increase with the increasing electric charge of  the black hole, which is qualitatively different from the extreme Kerr-Sen black hole case.

\section{Variation of escape probability with respect to position of particle source}\label{10013}
In the previous sections, we investigated the escape probability of the particle from the source near the Kerr-Sen black hole and explored the effect of the black hole parameters on it. We will now study the escape probability variation of the particle with respect to the position of the particle. As seen in above, there is no qualitative difference between the massless particle case and the massive particle case, we will now choose the massive particle case as an example to illustrate the characteristics of the escape probability variation in terms of the particle position.
 
We find that, as shown in the upper diagram of Fig. \ref{ep3}, outside the extreme Kerr-Sen black hole, for the massive particle, if $Q<0.82$, the escape probability first decreases and then increases with the increasing distance apart from the horizon; or else, the escape probability increases monotonically. The reason is that, with the increasing of the electric charge of the extreme Kerr-Sen black hole, the extreme points of the escape probabilities move nearer and nearer to the event horizons; and if $Q>0.82$, the extreme points are inside the event horizons. (We also find that, the massless particle case is similar to the massive particle case, and the critical electric charge becomes $Q=0.69$. We do not show the figure of this escape probability.)

In the non-extreme Kerr-Sen black hole case, as seen in the bottom diagram of Fig. \ref{ep3}, we can know that the escape probabilities of the particles first decrease and then increase monotonically with the increasing of the position of the particle source. There is a cusp point on each curve. This is quite different from the extreme Kerr-Sen black hole case.

\section{Conclusions and Discussions}\label{5411}
We considered the escape probabilities of the massless and massive particles produced by a particle source which is at rest in a LNRF on the equatorial plane  of the Kerr-Sen black hole. Supposing that the particles escape from the black hole to spatial infinity isotropically, we found that the escape probability of the particle decreases with the electric charge (or increases with the angular momentum) for the extreme Kerr-Sen black hole at the horizon limit but increases with the electric charge (or decreases with the angular momentum) for the non-extreme Kerr-Sen black hole near the horizon.  This illustrates that the angular momentum of the Kerr-Sen black hole plays a role in promoting (or suppressing) the escape probability of the particle. This implies that the rapidly rotating extreme Kerr-Sen black hole or the slowly rotating non-extreme Kerr-Sen black hole can be easier to be observed by a distant observer. In this regard, our results are quite different from those in  \cite{Ogasawara:2019mir,Igata:2019hkz}.

We also studied the escape probability variation of the particle with respect to the radial position of the particle source. We found that there is a critical electric charge for the extreme Kerr-Sen black hole, beyond which the escape probability of the massless or massive particle increases monotonically with the increasing distance apart from the event horizon, and below which the probability decreases first and then increases. In the non-extreme Kerr-Sen black hole background, we found that the escape probability of the particle first decreases  and then increases with the radial position of the particle source, which is quite against common sense. The results shown in Fig. \ref{ep3} indicate a fact, that is, it can be more difficult for a particle at a position beyond the horizon than for a particle at a position near the horizon of the Kerr-Sen black hole in certain conditions.

\section*{Acknowledgements}
Jie Jiang is supported by the National Natural Science Foundation of China (Grant Nos. 11775022 and 11873044). M. Zhang is supported by the Initial Research Foundation of Jiangxi Normal University with Grant No. 12020023.

\bibliography{Notes}

\end{document}